\documentstyle[prl,aps,preprint]{revtex}
\begin{document}

\title{Optical properties of a two-dimensional electron gas \\ at
even-denominator filling fractions} \author{Darren J.T. Leonard$^1$, Neil F.
Johnson$^1$, V. Nikos Nicopoulos$^1$ and F.J. Rodriguez$^2$} %
\address{$^1$ Department of Physics, Clarendon Laboratory, Oxford University,
Oxford OX1 3PU, England}
\address{$^2$ Departamento de Fisica, Universidad de
Los Andes, Bogota A.A. 4976,
Colombia}
\maketitle

\begin{abstract} The optical properties of an electron gas in a magnetic field at
filling fractions  $\nu={1\over 2m}$ ($m=1,2,3\dots$) are investigated using the
composite fermion picture. The response of the system to the
presence of valence-band holes is calculated. The shapes of the
emission spectra are found to differ qualitatively from
the well-known electron-hole results at zero magnetic field.
In particular, the asymmetry of the emission lineshape is found to be sensitive to the hole-composite fermion plane separation.
\end{abstract}

\newpage

\section{Introduction} Recent theoretical studies of the properties of a
two-dimensional electron gas at filling factor $\nu = {1\over 2m}$~\cite{HLR,AS}
have shown that the system of electrons at finite magnetic field can be
mapped on
to one of new quasi-particles, composite fermions (CFs), which do not
experience any magnetic field. An additional bonus of this mapping is that the
resulting CFs have weak interactions, even though the original electrons were
strongly interacting. These remarkable results are obtained by applying a
singular gauge transformation to the electron system; this transformation has
the effect of attaching $2m$ quanta of magnetic flux to each electron. At the
mean field level, the ground state of the system is now one of a Fermi liquid of
CFs. Studies have centred on the transport properties of the CF gas -- to our
knowledge no work has yet been published regarding its optical properties. This
is in spite of recent experiments which optically probe the
two-dimensional electron gas in the Fractional Quantum Hall regime around
$\nu = {1\over 2}$ (e.g. Refs. \cite{HARRIS} and \cite{HENRY}).

The purpose of this paper is to investigate the optical signature
of a CF gas at even denominator filling fractions $\nu = {1\over 2m}$.
The CFs in this paper are treated  at the mean-field level only. As a
consequence, the CFs are non-interacting particles \cite{HLR,AS}; this
simplifies the treatment of the CF-hole interaction since there will be no
dielectric screening by the CF gas. In the seminal paper of Halperin, Lee and
Read \cite{HLR}, the response of the CF gas to impurity potentials is
considered. The localized potential causes density fluctuations in the CF
gas and consequently a fluctuation in the magnetic field. In the
experimental heterostructures of Turberfield  and co-workers \cite{HARRIS},
however, the valence-band hole is delocalized, i.e. the corresponding hole
potential is not localized. Consequently we do not expect the effect of the
induced vector potential to be so important. The valence hole can therefore be
reasonably described as a weak (delocalized) probe with the interaction strength
taken as a parameter dependent on the plane separation $d$. Within the
mean-field
approximation, therefore, the problem  reduces to that of a gas of CFs, which
experience no magnetic field, interacting with valence holes which feel the full
$\nu={1\over 2m}$ magnetic field (see Fig. 1). This feature makes the problem
different from the case of electrons and holes interacting at zero magnetic
field. As we will show, this has an interesting effect on the shape of the
optical spectra, which depends on the competition between the effects due to the
dipole matrix elements and those due to the CF-hole interaction.

In accordance with the experimental GaAlAs heterostructure
systems containing delocalized holes \cite{HARRIS,HENRY}, our model
considers CFs and holes constrained to parallel planes separated by a distance
$d$. The CF-hole interaction depends on this separation and as a result the
optical properties are altered by a change of $d$. The treatment of many-body
optical effects presented here does not address some of the more subtle issues
concerning hole self-energies that have been discussed recently in connection
with Fermi-edge enhancement effects in the two-dimensional electron gas.
However, the simplifications employed here do allow a number of analytic results
to be obtained, thereby reducing the build-up of numerical
errors. The general qualitative features obtained here should also arise in a
fuller treatment. It is known that the hole self-energy diagrams contribute
greatly to the suppression of the Mahan exciton into a Fermi-edge singularity as
a result of the orthogonality catastrophe
\cite{MAHAN,RODRIGUEZ,SCHMITT,MND,RUCKEN}. This effect is crudely taken into
account here via the broadening of the hole spectral function from a
delta-function to a Lorentzian. The form employed here permits an {\em analytic}
calculation of the zero-order CF--dressed-hole Green function at finite
temperature; this function represents the building block for a
ladder-approximation treatment of CF-hole interactions.

The plan of the paper is as follows. In Sec. II, we review the relevant
single-particle results. Section III provides the formal expressions used in
the calculation of the optical properties; in particular, the dielectric
function (III.A), the dipole matrix elements (III.B) and the CF-hole Green
function (III.C and III.D). Specific optical spectra are discussed in Sec. IV
at half-filling, with Sec. V providing the conclusions.

\vskip\baselineskip

\section{Single Particle Properties}
\subsection{Composite Fermions}

The construction of the composite fermions (CFs)~\cite{HLR,AS,JAIN} from the
conduction band electrons at $\nu = {1\over 2m}$ is achieved by making a gauge
transformation which attaches $2m$ magnetic flux quanta to each electron. At the
mean field-level this cancels out the externally applied magnetic field and
screens completely the interaction between the particles. In the picture of CFs
presented by Jain~\cite{JAIN,READ}, the many-electron wavefunction of
interacting electrons in a magnetic field is constructed from the wavefunction
of non-interacting particles in a smaller magnetic field by attaching vortices
to each particle, thereby introducing zeroes into the wavefunction.
The vortices have the property of keeping the particles apart from each other,
thereby modelling the effect of the repulsive interaction, and also of
decreasing the filling factor. The net result is that the original interacting
electron system at  even denominator filling fractions is
mapped on to the non-interacting CF system at zero effective magnetic field
(c.f. Fig. 1).

At even denominator filling fractions the CF wavefunctions, normalised to an
area $L^2$,  are taken to be \begin{equation}\Phi ({\bf r}) = {1\over L}
e^{{\textstyle i {\bf K} \cdot {\bf r}}} \label{eq:CF wavefn} \end{equation}
with energy $E \equiv \hbar \omega({\bf K}) = {\hbar ^2 {\bf K}^2 \over
2m_{CF}^\star} + E_{gap}$\,; $\bf{r}$ and ${\bf K}$ are the two-dimensional position and momentum vectors respectively. CF creation and annilihation operators
$c^{\dag}_{{\bf K}}$ and $c_{{\bf K}}$ satisfy anticommutation relations
$\{c_{{\bf K}},c^{\dag}_{{\bf K^{\prime}}}\} = \delta_{{\bf K} ,{\bf
K^{\prime}}}$. Ignoring the fluctuations of the gauge field and assuming the CFs
form a Fermi liquid with a renormalized effective {\it optical} mass
$m^{\star}_{CF}$ , one may write the CF spectral function as $A_{CF}({\bf
K},\omega) = 2\pi \delta(\omega({\bf K}) - \omega)$. The CF chemical potential
must be determined by the fact that the system is at filling factor $\nu =
{1\over 2m}$. Since for spinless particles in two dimensions the number density
$n$ is related to the Fermi wave vector by $4\pi n = k^2_f$ , and related to the
filling factor by $\nu = 2\pi l^2_o n$, then at $\nu = {1\over 2m} \hbox{ one
sees that } k_f ={1\over l_0\sqrt{m}}$.

\subsection{Hole States}

The valence band holes (Fig. 1) are taken to have a single particle hamiltonian
\begin{equation} H = {({\bf P} - e{\bf A})^2 \over 2m^{\star}_h } \label{eq:VB
Hamiltonian} \ .
\end{equation} Choosing the Landau gauge with ${\bf A} = Bx{\bf {j}}$ leads to
states described by a Landau level index $n$ and a y-momentum $k$
\begin{equation} \Psi (x,y) = {1\over \sqrt{2^n n!\sqrt{\pi} l_0 L}}
e^{{\textstyle iky}} H_n\biggl({x-kl^2_0\over l_0} \biggr) e^{{\textstyle
-{1\over 2} \bigl({x-kl^2_0\over l_0}\bigr)^2}} \label{eq:VB wavefn}
\end{equation} which have energy $E = \hbar \omega_c (n+{1\over 2}) $ with
$\omega_c = {eB\over m_h^\star} $ and $l^2_0 = {\hbar \over m_h^\star
\omega_c}$, the functions $H_n(x)$ being the Hermite polynomials. An extremely
important many-body effect governing the optical properties is that the hole
states $|n,k\rangle$ can scatter from the low-energy CF excitations at the Fermi
level. This results in a self-energy which can be written as a Lorentzian
spectral function. One can crudely, but effectively, model this
effect by approximating the spectral function as \begin{equation} A_h (k,\omega)
= {2\gamma \over (\omega - {1\over 2} \omega_c)^2 + \gamma^2} \label{eq:hole
spectral fn} \end{equation} where  typically $\gamma \sim 10^{-2}\,\omega_c$.
The hole creation and annihilation operators satisfy the anticommutation
relations $\{d_{n,k_h},d^{\dag} _{m,l_h}\} = \delta_{n,m} \delta_{k_h,l_h}$. In
this work the Landau level index is always zero and hence can be dropped.

\vskip\baselineskip

\section{Optical Response} \subsection{Dielectric Function} The optical response
of a system to an external light field is characterized classically by the
dielectric function, which gives the refractive index and an absorption
coefficent~\cite{MAHAN,SCHMITT,HAUG}. Classically \begin{equation}
\epsilon(\omega)=1+\chi(\omega)
\end{equation} where \({\displaystyle \chi(\omega)}\) is the susceptibility in
the long-wavelength photon limit, and quantum mechanically \begin{equation}
\chi(\omega)=-{e^2\over \epsilon_0 V} R(\omega+i\delta) \qquad R(\omega+i\delta)
\equiv \langle\!\langle r(t);r(0) \rangle\!\rangle ^{irr} _{\omega + i\delta}
\label{eq:dielectric} \end{equation} Writing
\(\epsilon(\omega)=\epsilon_1(\omega)+i\epsilon_2(\omega)\), then the refractive
index \({\displaystyle n(\omega)=\sqrt{{1\over 2}
\biggl(\epsilon_1(\omega)+\bigl(\epsilon_1(\omega)^2+\epsilon_2(\omega)^2
\bigr)^{1\over 2}\biggr)}}\) and the absorption coefficient \({\displaystyle
A(\omega)={\omega\,\epsilon_2(\omega)\over c\,n(\omega)}}\). It is
common practice to take $A(\omega)\propto \epsilon_2(\omega)$ close to the band
gap, however this is an approximation since there can be an appreciable dependence of
$n(\omega)\) on \(\omega$ in this region. In order to calculate $A(\omega)$ one
must have exact details of the system studied such as the size and the dipole
matrix elements linking the valence and conduction band Bloch states.
Typically the actual function plotted is just
proportional to $ \epsilon_2(\omega)$. In the region of the spectrum where $\epsilon_2(\omega)$ is negative, one has net optical gain since $A(\omega)<0$; we label this as the emission spectrum. Similarly we label the region where $A(\omega)>0$ to be the absorption spectrum.  

\subsection{Dipole Matrix Elements} The dipole operator correlation function in
Eq.~(\ref{eq:dielectric}) is expanded in the CF-hole basis and reduces to
\begin{equation}R(\omega+i\delta) = \sum_{{\bf K},k,{\bf K^{\prime}},k^{\prime}}
\> r_{cv}^{\star} ({\bf K},k) r_{cv} ({\bf K}^{\prime},k^{\prime}) \biggl(
G(k,{\bf K};{\bf K}^{\prime},k^{\prime};\omega + i\delta) + G(k,{\bf K};{\bf
K}^{\prime},k^{\prime}; -\omega - i \delta) \biggr) \label{eq:dipole corrln}
\end{equation} where the two Green functions can be calculated using the usual
Matsubara formalism assuming quasi-equilibrium conditions for the number density
of particles. The CF states are labelled by the two-dimensional wavevector ${\bf
K}$, whereas the hole states are labelled by a wavevector $k$ in the
y-direction.
The interband matrix element in the envelope approximation for a transition from
the hole state $|\,0,k\rangle$ to the conduction band state $ |{\bf K}\rangle $
is given by \begin{equation} r_{cv} ({\bf K},k) = r_{cv} \delta_{-k,K_y} \sqrt
{{2 l_0 \sqrt \pi \over L}} S({\bf K}) \qquad S({\bf K}) \equiv e^{{\textstyle
iK_x K_y l^2_0 - {1\over 2} (K_x l_0)^2}}\label{eq:dipole} \end{equation} where
$r_{cv}$ is the matrix element of $\hat r$ between the valence and conduction
band Bloch states, assumed to be independent of ${\bf K}$. The novel feature of
the present matrix elements, as compared with those obtained
for the well-known situation of plane-wave valence and conduction states,
is that there is only conservation of momentum in the y-direction; the damping
of the matrix element in $K_x$ is exponential and leads to a damping in the
optical emission spectrum.

\subsection{CF-Hole Green Function} \subsubsection{Notation} The procedure to
determine the optical response is now to calculate the Green function
\begin{equation}G(k,{\bf K};{\bf K}^{\prime},k^{\prime};i\omega_n) \equiv
-\int_0^\beta d\tau e^{{\textstyle i \omega_n \tau}} \langle\!\langle T
d_{k}(\tau) c_{{\bf K}}(\tau) ;c^{\dag}_{{\bf K}^{\prime}}
d^{\dag}_{k^{\prime}}\rangle\!\rangle \label{eq:Two Part Gn fn}\end{equation} in
a perturbation expansion~\cite{MAHAN,SCHMITT}. Due to the presence of the
y-momentum-conserving terms in the dipole correlation function, the hole
wavevector $k = -K_y$, where $K_y$ is the CF y-wavevector. Thus the Green
functions are dependent on only four labels i.e. the four components of ${\bf K} \hbox{ and } {\bf
K^{\prime}}$. Hence the notation for the Green function to be calculated can be
simplified by defining \begin{equation}G({\bf K};{\bf K}^{\prime};\xi) \equiv
G(-K_y,{\bf K};{\bf K}^{\prime},-K^{\prime}_y;\xi) \qquad G({\bf K};\xi) \equiv
G({\bf K};{\bf K};\xi) . \label{eq:Gn fn A}\end{equation}
Defining \begin{equation}\Pi(\xi) = \sum_{{\bf K},{\bf K^{\prime}}}
S^{\star}({\bf K}) S({\bf K^{\prime}}) G({\bf K},{\bf K^{\prime}};\xi )
\label{eq:Pi A} \end{equation} one can write \begin{equation} A(\omega) \propto
- Im\  B(\omega) \qquad B(\omega) \equiv \Pi(\omega+i\delta) + \Pi(-\omega -
i\delta) . \label{eq:absorption coeff} \end{equation}

\subsubsection{Perturbation Expansion of $G({\bf K};{\bf K^{\prime}};i\omega)$}

The perturbation to be introduced is the interaction between CFs and holes.
CFs and holes have equal and opposite charges; a natural assumption for the
interaction form might be an unscreened Coulomb potential which
conserves y-momentum. The nature of the wavefunctions implies that the form of
the subsequent potential matrix elements is peculiar to the system,
\begin{equation}V({\bf K},k,k^\prime,{\bf K}^\prime) = -{e^2\over \epsilon_0
\epsilon_r} {e^{{\textstyle -|{\bf K}-{\bf K}^\prime|d}} \over |{\bf K}-{\bf
K}^\prime|} e^{{\textstyle -{1\over 4} l^2_0|{\bf K}-{\bf K}^\prime|^2 - {i\over
2} l^2_0 (K_x-K^\prime_x)(k+k^\prime)}} \label{eq:potential}\end{equation} where
$\epsilon_r \approx 13$ is the bulk GaAlAs dielectric constant, and $d$ is the
CF-hole separation in the direction perpendicular to the CF plane.
Unfortunately the complexity of these matrix elements makes subsequent analysis
 extremely difficult, even numerically. Our approach will therefore center
on the
use of a constant (i.e. momentum-independent) matrix element $V_0$ which allows
one to sum the diagrammatic expansion of the Green function in the ladder
diagram
approximation easily, and yet still retains the essential qualitative  features
of the true interaction \cite{MAHAN,RODRIGUEZ}. This produces a simple
analytical expression for the Green function in terms of integrations over ${\bf
K}$ of the zero-order CF-hole Green function. Poles of the Green function
indicating possible bound states, which leads to the observation of  excitonic
effects such as the Fermi-edge singularity, are then easy to find. In order
to choose an appropriate value for $V_0$, we set the momentum transfer in Eq.
(13) to be $|{\bf K} - {\bf K}^{\prime}| \sim k_f = {1\over
l_0\sqrt{m} }$ i.e. typical for scattering across the Fermi surface; we also
ignore the final oscillatory term in Eq. (13). The resulting matrix element
$V_0$ is then seen to scale with $d$ like $e^{-k_f d}$. The inter-plane
separation is therefore a parameter which can be used to scale the value of the
potential $V_0$. Spectra can be generated for different values of the scaling
factor $e^{-k_f d}$, hence the effect on the spectrum of different ratios
${d\over l_0\sqrt{m}}$ can be determined.

The diagrammatic expansion of $G({\bf K};{\bf K^{\prime}};i\omega)$
(see Fig.\,2) in this approximation becomes  \begin{equation}G({\bf K};{\bf
K^{\prime}};i\omega) = G_0({\bf K};i\omega) \delta_{{\bf K},{\bf K^{\prime}}} +
{G_0({\bf K};i\omega)
\; V_0 \; G_0({\bf K^{\prime}};i\omega)\over 1 - V_0 \; \sum_{\bf K^{\prime
\prime}} G_0({\bf K^{\prime \prime}};i\omega)} \label{eq:Gn fn
expansion}\end{equation} where the zero-order CF-hole propagator is $G_0({\bf
K};i\omega) .$ Hence from Eq.~(\ref{eq:Pi A}) \begin{equation} \Pi(i\omega)
=
\sum_{{\bf K}} |S({\bf K})|^2 G_0({\bf K};i\omega) + {V_0 \; \sum_{{\bf K}}
S^{\star}({\bf K}) G_0({\bf K};i\omega) \; \sum_{{\bf K}^{\prime}} S({\bf
K}^{\prime}) G_0({\bf K}^{\prime};i\omega)\over 1 - V_0 \; \sum_{{\bf K^{\prime
\prime}}} G_0({\bf K^{\prime \prime}};i\omega)}\label{eq:Pi
expansion}\end{equation} is the basis of the calculation of the optical
response. Analytically continuing $i\omega \rightarrow \omega + i\delta $ and $
-\omega -i\delta$ generates the correct functions $\Pi(\omega + i\delta)$ and
$\Pi(-\omega - i\delta)$ from which $B(\omega)$ results. However it is only
necessary to calculate $\Pi(\omega + i\delta)$  which is the `resonant' term,
not  $\Pi(-\omega - i\delta)$, for the study of the optical
response \cite{SCHMITT}. Since the matrix elements are an important quantity in
determing the shape of the spectra, it will also be interesting to examine the
spectra when the  matrix elements are constant over ${\bf K}$, i.e. $S({\bf K})
= 1$. This results in
\begin{equation} \Pi^{\prime}(i\omega) =
\sum_{{\bf K}} G_0({\bf K};i\omega) + {V_0 \; \sum_{{\bf K}} G_0({\bf
K};i\omega) \; \sum_{{\bf K}^{\prime}} G_0({\bf K}^{\prime};i\omega)
\over 1 - V_0 \; \sum_{{\bf K^{\prime
\prime}}} G_0({\bf K^{\prime \prime}};i\omega)}\label{eq:Pi2
expansion}\end{equation}

\subsection{Zero-Order CF-Hole Propagator} The zero-order CF-hole progagator
$G_0({\bf K},i\omega)$ can be written in the spectral
representation \cite{MAHAN,RODRIGUEZ,SCHMITT} as
\begin{equation}G_0({\bf K};i\omega_n) = \int_{-\infty}^{\infty} {d\omega_1\over
2\pi} \int_{-\infty}^{\infty} {d\omega_2\over 2\pi} {A_{CF} ({\bf K},\omega_1)
A_h (-K_y,\omega_2) [1 - f(\omega_1-{\mu_{CF}\over \hbar}) -
f(\omega_2-{\mu_h\over \hbar})]
\over i\omega_n - \hbar\omega_1 - \hbar\omega_2} .\label{eq:Gn
spectral}\end{equation} The function ${\displaystyle f(\omega) =
(e^{{\textstyle \beta \hbar \omega}} +1)^{-1}}$ is a Fermi-Dirac distribution
factor for the number of particles at a given energy; $\mu_{CF}$ and $\mu_h$
are the chemical potentials for the CFs and holes respectively. Making a change
of variables to $x={1\over 2} {\bf K}^2 l^2_0$ and $\Omega=
\omega-\omega_g-{1\over 2}\omega_c$ and with  $\mu_h = \hbar({1\over
2}\omega_c-\omega_\Delta)$, we obtain \begin{eqnarray}
\hbar G_0(x,\Omega &+& i\delta) [ (\Omega -\omega_c^\star x)^2+
\gamma^2 ]
= [1-f(\omega_c^\star(x-{\textstyle{1\over 2m}}))-f(\omega_\Delta+i\gamma)]
(\Omega-\omega_c^\star x)
  \\ \nonumber
 &-&  i\gamma [{\textstyle{1\over 2}} - f(\omega_c^\star(x-{\textstyle{1\over
2m}})) +  f(\omega_\Delta+i\gamma)  - f(\Omega-\omega_c^\star x +
\omega_\Delta)] \\ \nonumber
&+& {\gamma \over \pi} {\textstyle Re} [\psi({\textstyle  {1\over
2}-{i\beta\hbar\over 2\pi} (\Omega-\omega_c^\star x + \omega_\Delta)})] \\
\nonumber &+& {i\over 2\pi} \bigl[(\Omega-\omega_c^\star x+i\gamma)
\psi({\textstyle {1\over 2}- {\beta\hbar\over 2\pi}(\gamma-i\omega_\Delta)}) -
(\Omega-\omega_c^\star  x  -i\gamma)\psi({\textstyle {1\over 2}+
{\beta\hbar\over 2\pi}(\gamma+i\omega_\Delta)})\bigr]
\label{eq:Gn0}
\end{eqnarray}
where the Digamma function $\psi(z)$ is defined as
\begin{equation}
\psi(z) \equiv -C -\sum^\infty_{n=0}
\biggl( {1\over z+n} - {1\over 1+n} \biggr)
\end{equation}
with {\em C} being Euler's constant.
In terms of $G_0(x,\Omega+i\delta)$ and
$W = -{V_0\over 2\pi \hbar\omega_c l^2_0}$, which is the ratio of the Coulomb
energy to the kinetic cyclotron energy, we obtain
\begin{eqnarray}&\Pi&(\Omega + i\delta) =\\ \nonumber
& &{1\over 2\pi \hbar\omega_c l^2_0} \;
\Biggl[\int_0^{\phi} dx \, e^{-x} I_0(x) \, \hbar\omega_c G_0(x,\Omega +
i\delta) - {W\biggl( \int_0^{\phi} dx \, e^{-{1\over 2} x}
J_0\biggl({\sqrt{3}\over 2} x\biggr) \, \hbar\omega_c G_0(x,\Omega + i\delta)
\biggr)^2 \over 1 + W \int_0^{\phi} dx
\, \hbar\omega_c G_0(x,\Omega + i\delta)}\Biggr]  \label{eq:Pi
complete}\end{eqnarray} where $\phi$ is a cut-off determined by the size of the
Brillouin Zone associated with the bulk semiconductor environment. The Bessel
functions $I_0(x)
\hbox{ and } J_0(x)$ are defined as
\begin{equation} I_0(x) \equiv
\sum^\infty_{n=0} {1\over (n!)^2} \biggl({x\over 2}\biggr)^{2n} \qquad J_0(x)
\equiv \sum^\infty_{n=0} {(-1)^n\over (n!)^2} \biggl({x\over 2}\biggr)^{2n}\ \ .
\end{equation} We note that we have reduced the problem of calculating
$\Pi(\Omega+i\delta)$ to a relatively straightforward one-dimensional
integral over $x$. For the
case of constant matrix element $S({\bf K})=1$ the corresponding result is
\begin{eqnarray}\Pi^\prime(\Omega + i\delta)   ={1\over 2\pi \hbar\omega_c
l^2_0} \; \Biggl[{\int_0^{\phi} dx \, \hbar\omega_c G_0(x,\Omega + i\delta)
\over  1 + W \int_0^{\phi} dx\, \hbar\omega_c G_0(x,\Omega + i\delta)}\Biggr]
\label{eq:Piprime complete}\end{eqnarray}

\vskip\baselineskip

\section{Results at Half Filling} In this section the imaginary part of the
dielectric function is calculated at $\nu=\frac{1}{2}$ for  parameters taken in
accordance with the GaAlAs heterostructure system studied in Ref.
\cite{HARRIS}. The areal density of electrons (CFs) is $10^{15}{\rm m}^{-2}$,
the hole effective mass is taken as $0.45m_e$, and an electron and hole
temperature of $0.1 {\rm K}$ is used. One further parameter is needed which is
the CF effective mass. On the basis of the  expectation that at $\nu={1\over 2}$
the emission spectrum should occur over the Fermi  surface, we can equate the
width of the peak observed in experiments in Ref.  \cite{HARRIS} to the Fermi
energy $E_f={1\over 2}\hbar\omega_c^\star$; the particular data used implies $m_{CF}^\star=0.70m_e$ with $\omega_c^\star = 0.64\omega_c$. The chemical potential of the holes is  set so
that the hole number density is small compared to the number of CFs, and yet
there is still a measurable emission spectrum; we use $\omega_\Delta =
0.02\omega_c$ which gives ${n_h\over n_{CF}}=0.15$. The spectra are plotted against a photon energy $\hbar\Omega$ in units of the Fermi energy.

Enhancements of the optical properties of the CFs occur as a result of the
attractive interaction between the CFs and the holes, which try to form
CF-excitons.    It is clear from Eq.~(\ref{eq:Pi complete}) that there is
an excitonic enhancement of the  optical response when
\begin{equation}1 + W \int_0^{\phi} dx\, \hbar\omega_c Re\;G_0(x,\Omega + i\delta) =
0\ \ .
\label{eq:condition}\end{equation} The frequencies for which this is true are
the Mahan exciton states \cite{MAHAN}, these being the equivalent of the
two-particle exciton states when there is a large number density of CFs. These
excitonic states only form when the potential is sufficiently strong to bind up
CFs and holes. Since the strength of the potential is determined by the
separation of the CF and hole planes $d$, then the separation is crucial in
deciding the strength of  enhancement and hence the lineshape of the
optical spectra. Figure 3 shows the real and  imaginary parts of
$\int_0^{\phi} dx\, \hbar\omega_c  G_0(x,\Omega + i\delta)$; it also shows
plots of $-{1\over W}$ (dashed lines) which arise from different $d$ as
indicated, and condition (\ref{eq:condition}) is satisfied when these
dashed lines cross the solid curve, with the Mahan exciton frequencies being the
points of intersection. As can be seen from the figure,  the potentials used
here are {\em not} strong enough to create bound Mahan excitons; however it is
clear from the plots of the imaginary part of the dielectric function that the
interaction enhances the spectra at the Fermi surface due to the increasing
overlap of CFs and holes, even if the potential is not strong enough to bind
them up.

Figure 4 shows the emission part of the spectrum for CFs and holes separated by
various inter-plane distances $d$. We note that the dipole matrix
elements will, in principle, also be dependent on the overlap of the parts
of the
CF and hole wavefunctions which lie in the `z' direction, i.e. perpendicular to
the planes; however these will mostly affect the {\em intensities} of the
spectra
but {\em not the lineshapes}. We are only  interested in the variation of
the lineshape with the separation $d$.  The curve showing emission when there is
no interaction, exhibits an intrinsic asymmetry in the lineshape with a
{\em left bias}; this results purely from the dipole matrix elements and
hence $S({\bf K})$.  As a comparison, Fig.~5 shows the emission
when this factor is taken to be constant, i.e. from  Eq.~(\ref{eq:Piprime
complete}). Figures 4 and 5 illustrate the  difference that the
CF-hole matrix element has on the lineshape. One can also make  comparisons
between the intensities for lines with the same $d$; it is noticeable that
the emission intensity is reduced by the inclusion of the correct factor $S({\bf
K})$.  Switching on the interaction changes the asymmetry from {\em left}
to {\em
right} as  $d$ is decreased; this results from enhancement appearing at the
Fermi
surface which is essentially of logarithmic-divergence
character~\cite{MAHAN}. The interaction potential  has not been allowed to be
strong enough to cause the Mahan exciton to bind, i.e.
Eq.~(\ref{eq:condition}) is never satisfied. This is because such a
singularity is  always suppressed in practice by the orthogonality
catastrophe~\cite{MAHAN,RODRIGUEZ,MND} into a straightforward Fermi-edge
enhancement.
The results shown in Fig.~4 can also be applied to other even-denominator
fractions $\nu=\frac{1}{2m}$, after a simple rescaling of $\omega_c^*$.
Interestingly, the predicted asymmetry in the emission lineshape (see Fig. 4)
seems consistent with recent photoemission experiments \cite{HENRY}, where
either a left or right-biased lineshape can be observed according to the
specific heterostructure sample and fraction $\frac{1}{2m}$ being studied, i.e.
according to the magnitude of $\frac{d}{l_0}$.  A detailed comparison with this
experimental data will be presented in a future publication \cite{DARREN}.

Figures 6 and 7 show the absorption part of the spectra. As with Figs.~4
and 5, Fig.~6 is  calculated including the correct factor $S({\bf
K})$ while Fig.~7 has this factor set to unity. The effect
of the ${\bf K}$-dependence is not as clear here as  it is in emission, but it
is still evident. Consider the spectra when the interaction is  turned off;
in Fig.~7 the curve is constant above about $1.2E_f$, being consistent
with the step function asssociated with absorption at zero temperature and with
no hole broadening~\cite{MAHAN}. However there is a decay at higher energies in
Fig.~6 associated with the factor $e^{-x} I_0(x)$ in Eq.~(\ref{eq:Pi
complete}) resulting directly from the matrix element factor  $S({\bf K})$.
When the interaction is turned on, one observes the enhancement at the
Fermi-edge  resulting from the tendency to form Mahan excitons, with the spectra
in Fig.~6 being smaller in amplitude than in Fig.~7, again as a result
of the
${\bf K}$-dependence of the matrix element.

\vskip\baselineskip
In this paper the effect of CF-CF interactions is not considered. These
interactions
would tend to screen the true CF-hole interaction in a non-trivial
fashion, as a result of the induced vector potential which accompanies
density fluctuations caused by the presence of the hole. However, the
inclusion of CF-CF interactions should not significantly alter the
qualitative lineshapes presented here for two reasons. First, the
CF-CF interaction cancels within the mean-field approximation in the
absence of a localized external potential. Introducing the hole will
therefore, at most, introduce a residual CF-CF interaction which is
smaller than the electron-electron interaction in the more familiar
electron-hole problem. (In the electron-hole problem at $B=0$ it is
known that electron-electron interactions change the functional form
of the lineshape at the edges, but not the overall shape). Second, our
model uses a constant effective CF-hole interaction -- as a first
approximation, this constant potential can be thought of as including 
CF-CF screening effects by representing an effective, screened CF-hole
interaction. 

The subtleties of a modified form of the screened CF-hole interaction
due to CF-CF interactions should not, therefore, affect the primary
result of this 
paper, which is that there will be some degree of Fermi-edge
enhancement. Instead, the CF-CF interactions are likely to affect the
detailed lineshape at the Fermi edge ~\cite{MAHAN,MND}. 
The effect of CF-CF interactions on the effective mass of the CF would
be such as to cause a 
divergence at the Fermi surface. This will influence primarily the
edge behavior of 
the spectra. The broadening of the spectra at the Fermi energy
essentially 
depends on the ratio of the CF effective mass to the hole effective 
mass~\cite{MAHAN,RODRIGUEZ,MND}.  Since 
the latter is infinite whilst the former is finite except at the Fermi
energy, there
is likely to be little effect as a result of the infinite CF mass at
the Fermi edge. The optical response probes all CF states, therefore 
the use of a single {\em optical} effective mass to describe all the
CFs is reasonable. 
This mass $m^\star_{CF}$ is chosen to fit the experimental 
data~\cite{HARRIS} equating the Fermi energy ${\hbar^2 k_f^2\over
2m^\star_{CF}}$ to 
the width of the line in emission. We emphasize that we found its
fitted value to be of a similar order of magnitude to CF masses
obtained using transport measurements. 
This paper also does not address the question of how the  
hole Green function is renormalized by the interactions with the CFs.
This is an 
interesting, but difficult, question in itself; the fact that the hole
is mobile, 
i.e. it is not localized as an impurity state, precludes the use of
the
technique introduced by Nozi\`eres and de Dominicis\cite{MND}. These
authors approach considered the only 
effect of the hole as being the introduction of a static perturbation
to the CF gas for a
finite time interval, thereby decoupling the problem  into 
scattering of a single CF from an impurity potential, plus the
renormalization of  the 
(single state) hole Green function. The consequence of renormalizing
the localised 
hole is to broaden it's spectral function, thereby leading to a
suppression 
of the Mahan divergence in the optical response. The  inclusion of a
Lorentzian broadening for a mobile hole only addresses the fact that
physically 
there should be a similar suppression of the Mahan exciton due to the
hole 
shake-up of the CF sea. Assuming the hole spectral function to be
Lorenztian 
then allows one to perform some of the integrals analytically, thereby
producing 
simple closed forms for the optical response in terms of
one-dimensional integrals.
A more complete calculation of the hole renormalization could be
performed in the situation where the hole state were bound to an
impurity; this would allow one to 
determine the effect of the orthogonality catastrophe on the spectrum
taking
into account the dielectric properties of the CF gas.  However, our
focus in this paper was to consider the the class of experiments in
heterojunctions with (at least) nominally  delocalized holes.

\vskip\baselineskip

\section{Conclusion} This paper has calculated the optical signature of a
composite fermion (CF)-hole system at even-denominator filling fractions
$\nu=\frac{1}{2m}$. The shapes of the
emission and absorption spectra have been found to differ from  the well-known electron-hole results at zero magnetic
field.
The results suggest that emission lineshapes measured in photoluminescence
experiments on GaAlAs heterostructures at $\nu={1\over 2m}$ can be used
to extract information concerning the electron-hole plane separation.
Future work will focus upon including effects of gauge fluctuations (i.e.
effective CF-CF interactions) into the formalism; of particular interest is the
question of how such gauge fluctuations will affect the optical
spectra.

\vskip\baselineskip
\vskip\baselineskip

\noindent {\bf Acknowledgements}
We acknowledge the financial support of EPSRC through a Studentship (D.J.T.L),
EPSRC grant No. GR/K 15619 (N.F.J. and V.N.N.) and COLCIENCIAS 
grant No. 1204-05-264-94 (F.J.R.). We
would
like to thank Andrew Turberfield for very useful discussions. We are also
grateful to Robin Nicholas, Henry Cheng, Taco Portengen and Luis Quiroga. We
thank Henry Cheng for showing us his experimental data prior to publication.

\newpage \centerline{\bf Figure Captions}

\bigskip

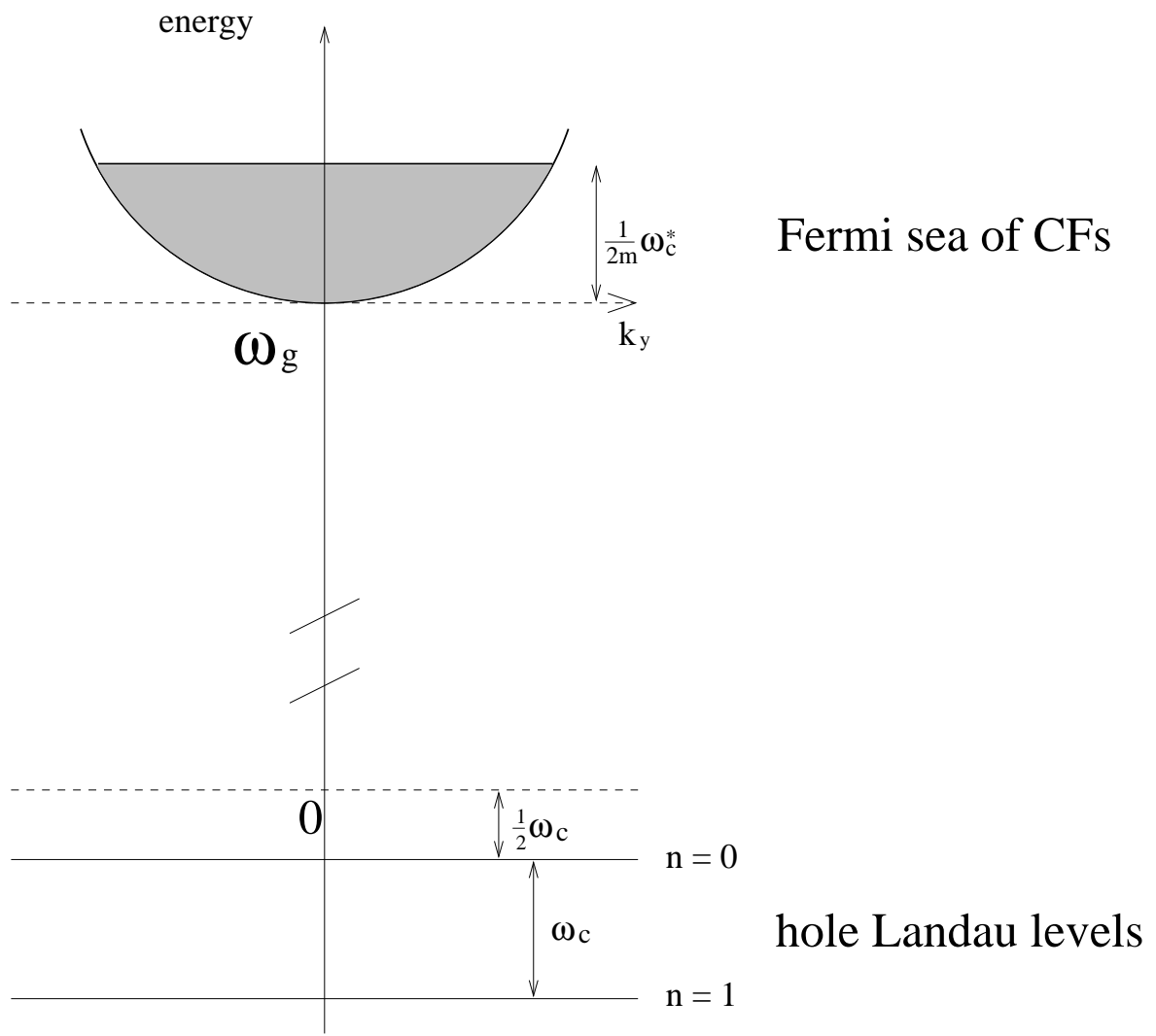
\noindent Figure 1: Effective composite fermion (CF) and hole band structure
diagram.

\bigskip

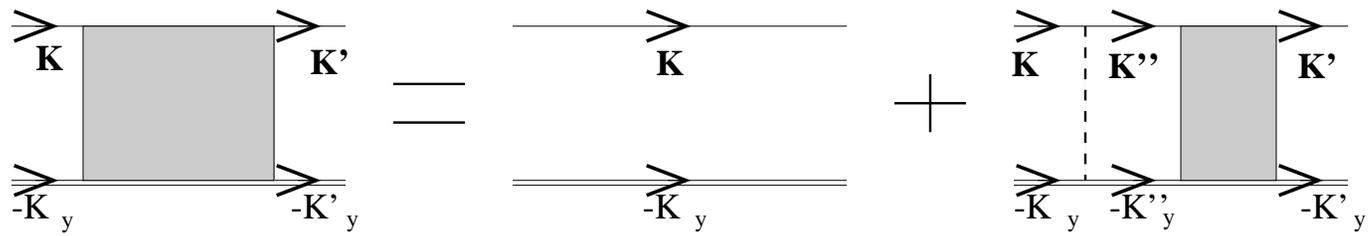
\noindent Figure 2: Diagrammatic expansion of $G({\bf K};{\bf
K^{\prime}};i\omega)$  in the ladder approximation. The diagrams are closed by the dipole matrix elements, which conserve y-momentum; since this is just the photon momentum, the only diagrams required are those where the total y-momentum is zero.

\bigskip

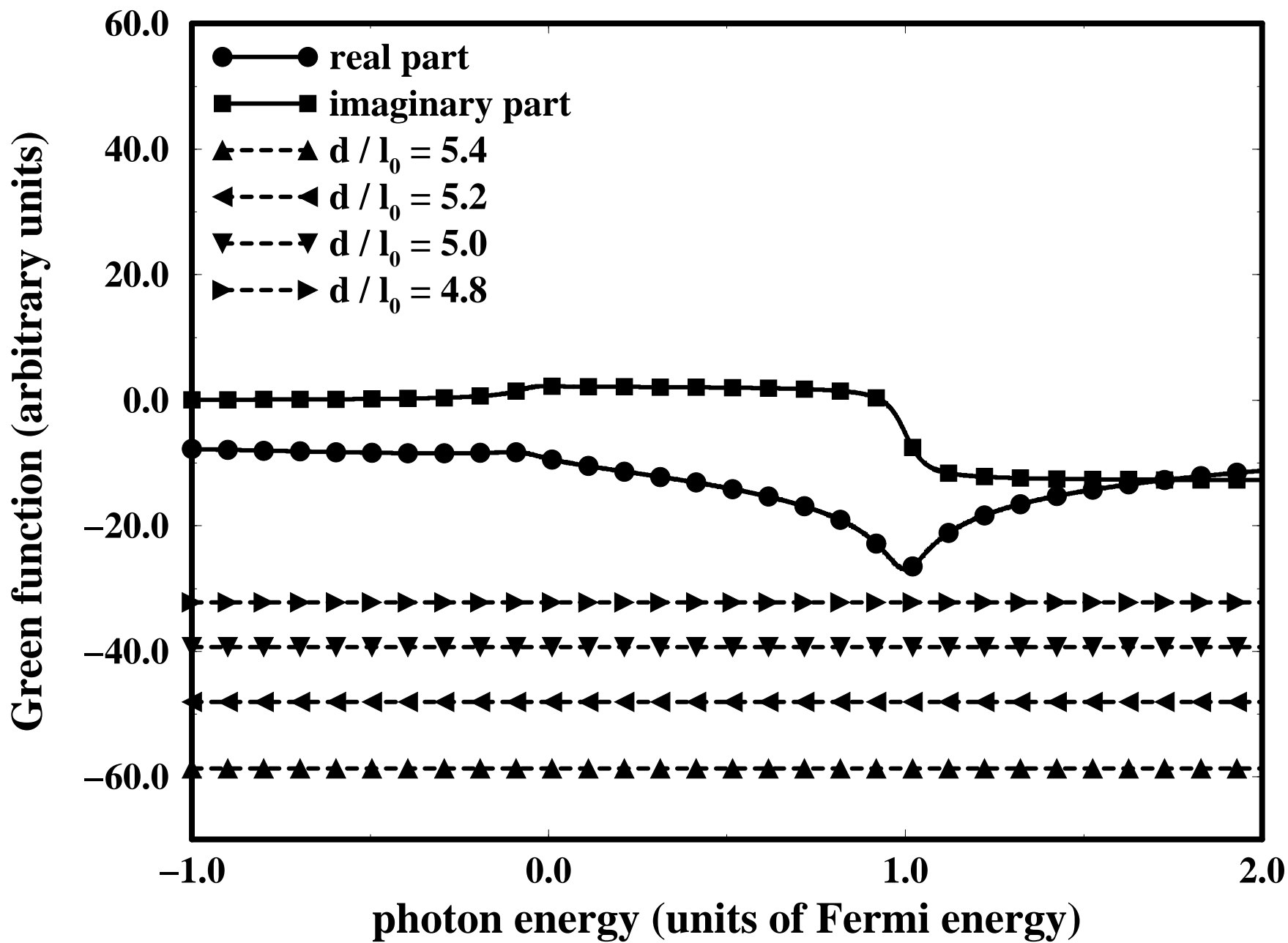
\noindent Figure 3: Real and imaginary parts of the integrated CF-hole Green
function (solid lines) together with plots of the inverse of the CF-hole
interaction strengths
$-{1\over W}$ for different CF-hole plane separations $d$ (dashed lines). The
magnetic  length $l_0$ is typically $89$A at $B=8T$ with $\nu={1\over 2}$ and
$n_e=10^{15}m^{-2}$. The Fermi energy $E_f=0.7meV$.

\bigskip

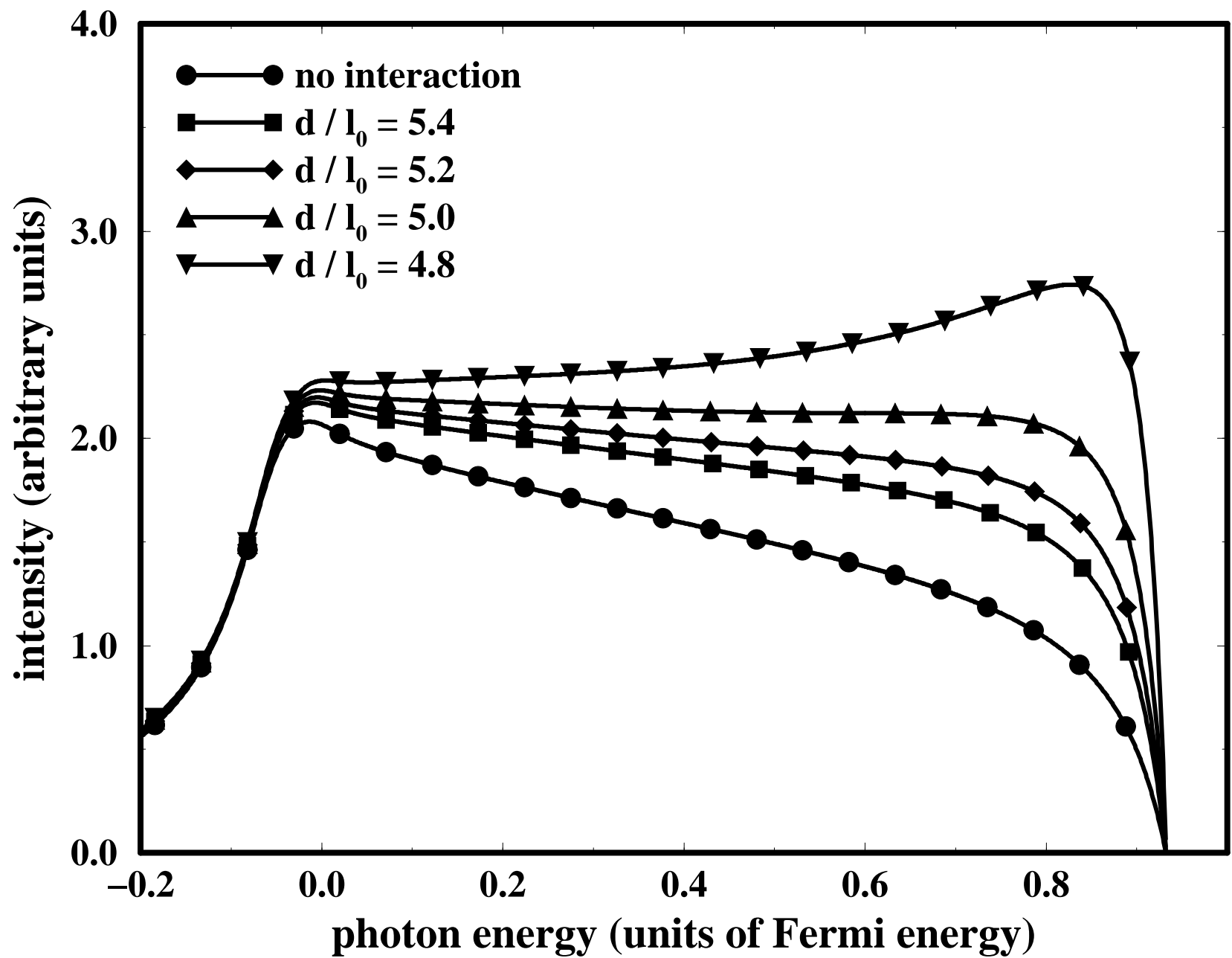
\noindent Figure 4: Emission  spectrum (i.e. $A(\omega)<0$) obtained from
the imaginary part of the dielectric function, plotted with different plane
separations $d$ and with the full {\bf K}-dependent matrix elements. Input
parameters correspond to GaAlAs heterostructure. The magnetic length $l_0$ is
typically $89$A at $B=8T$ with $\nu={1\over 2}$ and $n_e=10^{15}m^{-2}$. The Fermi energy $E_f=0.7meV$.

\bigskip

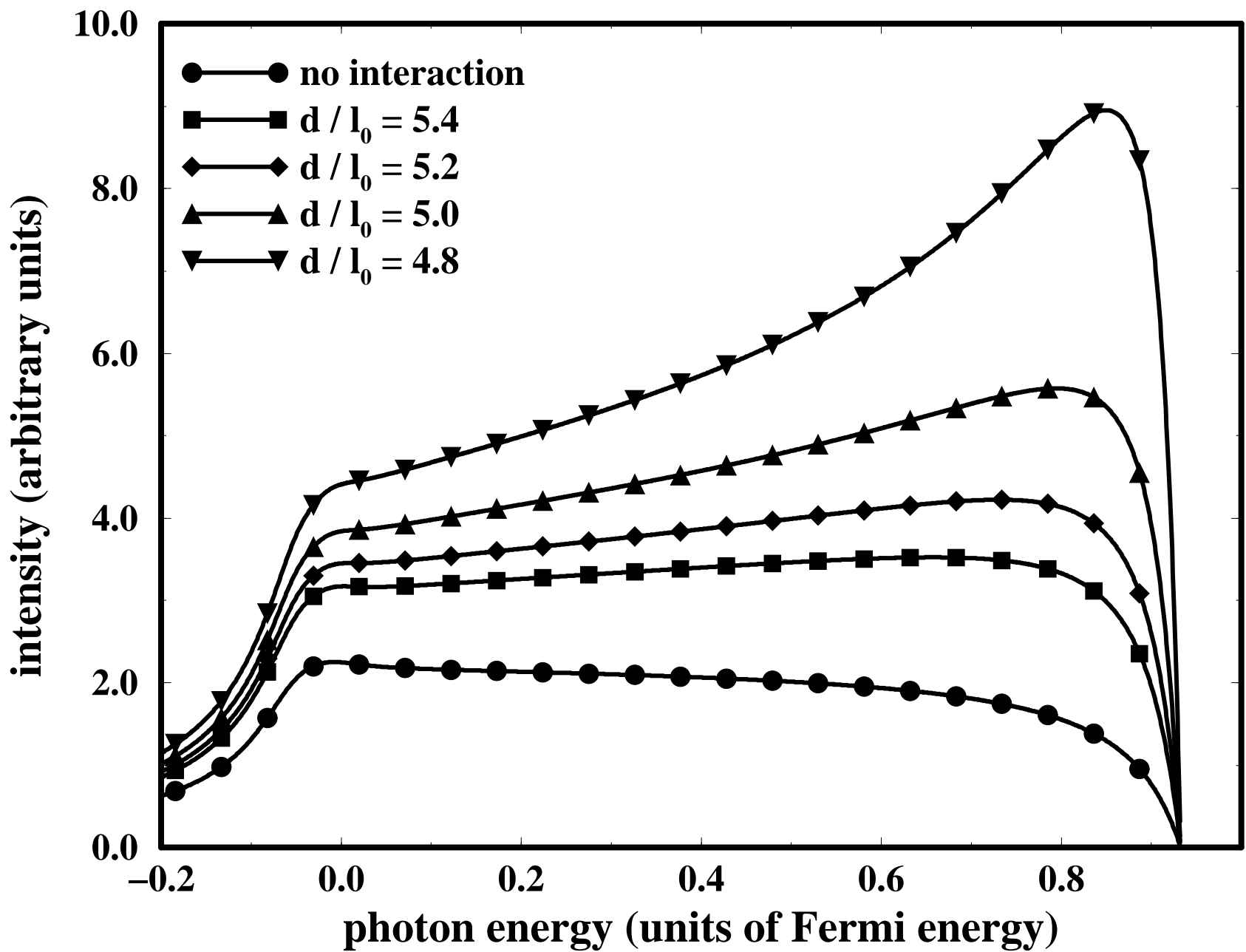
\noindent Figure 5: Emission  spectrum (i.e. $A(\omega)<0$) obtained from the imaginary part of
the dielectric
function, plotted with different plane separations $d$ as in Fig. 4, but {\em
without} {\bf K}-dependent matrix elements. Input parameters as in Fig. 4.

\bigskip

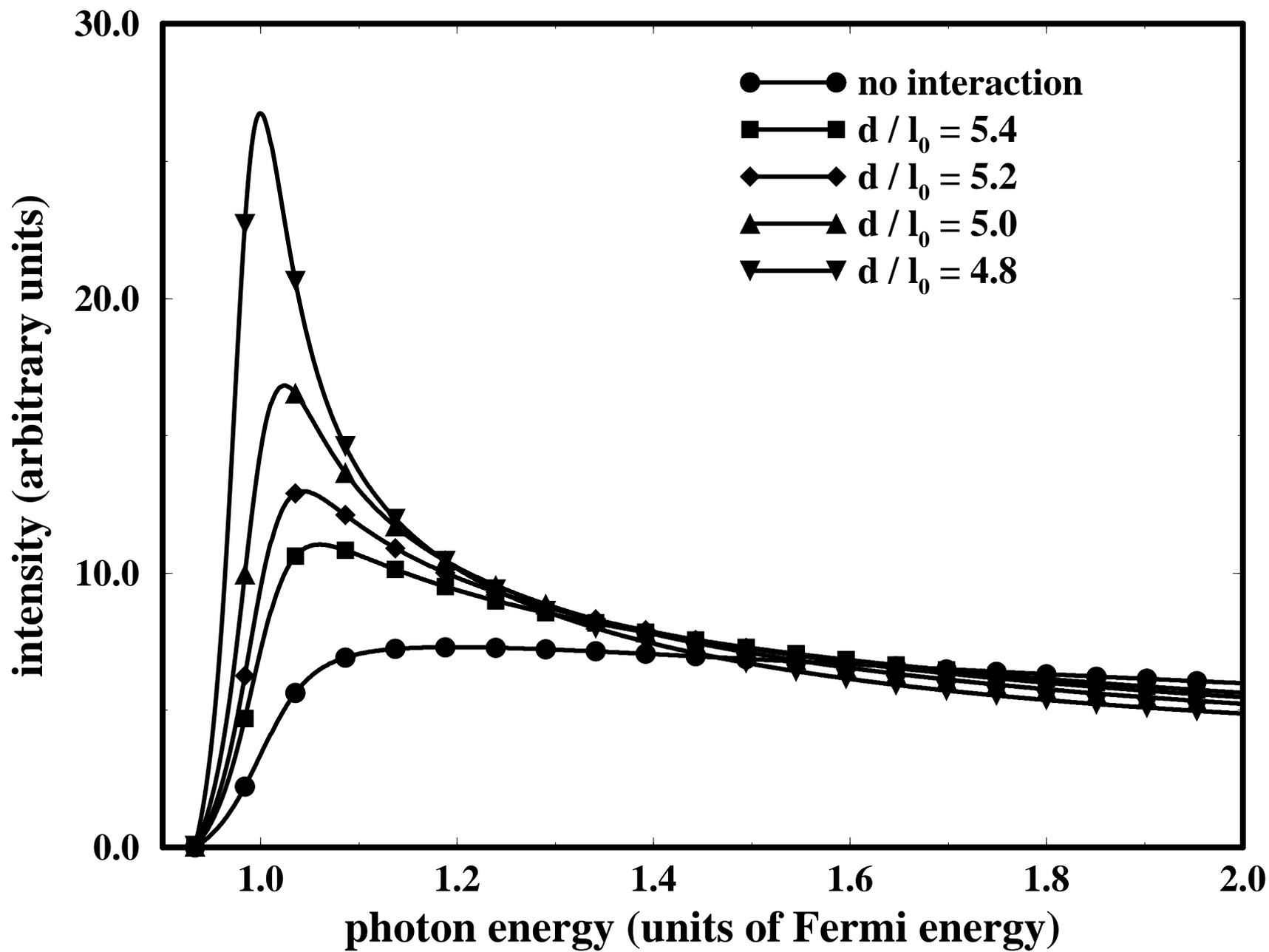
\noindent Figure 6: Absorption  spectrum (i.e. $A(\omega)>0$) obtained from the imaginary part of
the dielectric
function, plotted with different plane separations $d$ and with full {\bf
K}-dependent
matrix elements. Input parameters
correspond to Fig. 4.

\bigskip

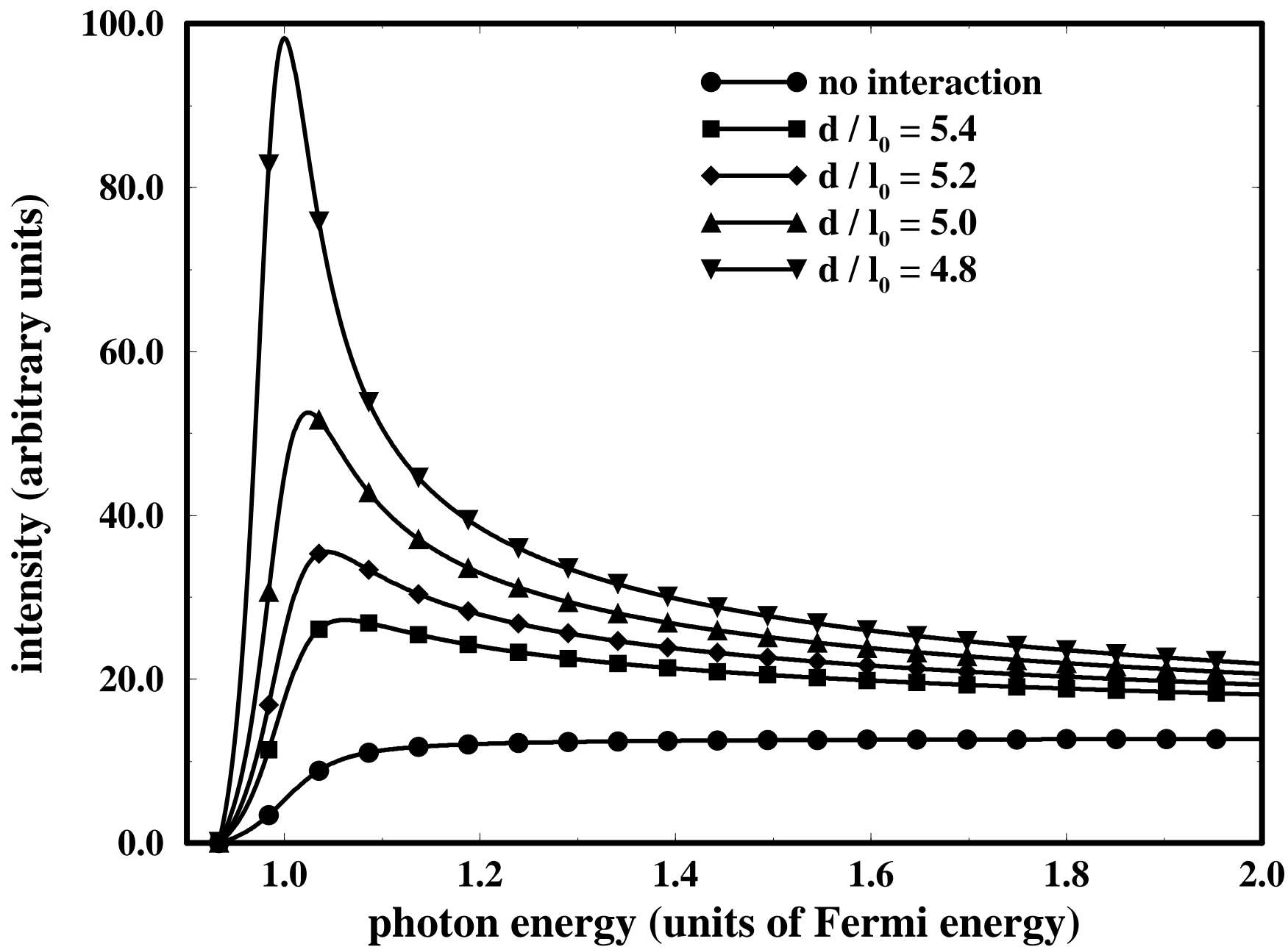
\noindent Figure 7: Absorption  spectrum (i.e. $A(\omega)>0$) obtained from the imaginary part of
the dielectric
function, plotted with different plane separations $d$ as in Fig. 6 but {\em
without} {\bf K}-dependent matrix elements. Input parameters
correspond to Fig. 4.

\end{document}